\documentclass[runningheads,a4paper]{llncs}
\usepackage[utf8]{inputenc}
\usepackage{amssymb}
\usepackage{url}
\usepackage{hyperref}
\usepackage{todonotes}
\usepackage{comment}
\usepackage{pifont}
\usepackage{amssymb}
\usepackage{graphicx}
\graphicspath{ {images/} }
\usepackage{amsmath}
\usepackage{algorithm2e}
\usepackage{listings}

\usepackage{url}
  
\newcommand{\keywords}[1]{\par\addvspace\baselineskip\noindent\keywordname\enspace\ignorespaces#1}

\usepackage{spverbatim}
\usepackage{fancyvrb}
\fvset{fontsize=\footnotesize}

\usepackage{paralist}

\newcommand{\prop}[1]{{\verb|#1|}}

\newcommand{\wikidata}{{Wikidata}\xspace}
\newcommand{\dbpedia}{{DBpedia}\xspace}
\newcommand{\DW}{{\scshape DBw}\xspace}
\newcommand{\rdf}{{RDF}\xspace}

\begin{document}

\mainmatter  

\title{Wikidata through the Eyes of DBpedia}

\titlerunning{Wikidata through the Eyes of DBpedia}

\author{Ali Ismayilov\inst{1} \and Dimitris Kontokostas\inst{2} \and S\"oren Auer\inst{1} \and Jens Lehmann\inst{2} \and Sebastian Hellmann\inst{2}}

\authorrunning{Ismayilov et al.}

\institute{
University of Bonn, Enterprise Information Systems and Fraunhofer IAIS
\\\email{s6alisma@uni-bonn.de|auer@cs.uni-bonn.de}
 \and Universit\"at Leipzig, Institut f\"ur Informatik, AKSW
\\\email{\{lastname\}@informatik.uni-leipzig.de}
}

%
%

\toctitle{Lecture Notes in Computer Science}
\tocauthor{Authors' Instructions}
\maketitle

\begin{abstract}
\dbpedia is one of the first and most prominent nodes of the Linked Open Data cloud.
It provides structured data for more than 100 Wikipedia language editions as well as Wikimedia Commons, has a mature ontology and a stable and thorough Linked Data publishing lifecycle.
\wikidata, on the other hand, has recently emerged as a user curated source for structured information which is included in Wikipedia.
In this paper, we present how \wikidata is incorporated in the \dbpedia eco-system.
Enriching \dbpedia with structured information from \wikidata provides added value for a number of usage scenarios. 
We outline those scenarios and describe the structure and conversion process of the \dbpedia \wikidata dataset.
\end{abstract}

\vspace{-1em}
\section{Introduction}
\vspace{-0.5em}
\dbpedia is one of the first and most prominent nodes of the Linked Open Data cloud.
It provides structured data for more than 100 Wikipedia language editions as well as Wikimedia Commons, has a mature ontology and a stable and thorough Linked Data publishing lifecycle.
\wikidata has recently emerged as a user curated source for structured information which is included in Wikipedia.

\dbpedia uses human-readable Wikipedia article identifiers to create IRIs for concepts in each  Wikipedia language edition and uses \rdf and Named Graphs as its original data model. 
\wikidata on the other hand uses language-independent numeric identifiers and developed its own data model, which provides better means for capturing provenance information. 
The multilingual \dbpedia ontology, organizes the extracted data and integrates the different language editions while \wikidata is rather schemaless, providing only simple templates and attribute recommendations.
All \dbpedia data is extracted from Wikipedia and Wikipedia authors thus unconciously also curate the \dbpedia knowledge base.
\wikidata on the other hand has its own data curation interface, which is also based on the MediaWiki framework.
\dbpedia publishes a number of datasets for each language edition in a number of Linked Data ways, including datasets dumps, dereferencable resources and SPARQL endpoints.
While \dbpedia covers a very large share of Wikipedia at the expense of partially reduced quality, \wikidata covers a significantly smaller share, 
but due to the manual curation with higher quality and provenance information.
As a result of this complementarity, aligning both efforts in a loosely coupled way would render a number of benefits for users.
\wikidata would be better integrated into the network of Linked Open Datasets and Linked Data aware users had a coherent way to access \wikidata and \dbpedia data.
Applications and use cases have more options for choosing the right balance between coverage and quality.

In this article we describe the integration of \wikidata into the \dbpedia Data Stack.
People are not used to the currently very evolving \wikidata schema, while \dbpedia has a relatively stable and commonly used ontology. 
As a result, with the \dbpedia{}  \wikidata (\DW) dataset can be queried with the same queries that are used with \dbpedia.

\vspace{-1em}
\section {Background}
\label{sec:background}
\vspace{-0.5em}
\paragraph{\wikidata}
\cite{wikidataMain} is community-created knowledge base to manage factual information of Wikipedia and its sister projects operated by the Wikimedia Foundation. 
As of March 2015, \wikidata contains more than 17.4 million items and 58 million statements.
The growth of \wikidata attracted researchers in Semantic Web technologies. 
In 2014, an \rdf export of \wikidata was introduced~\cite{wikidataRDF} and recently a few SPARQL endpoints were made available as external contributions. 
\wikidata is a collection of entity pages. 
There are three types of entity pages: items, property and query. Every item page contains labels, short description, aliases, statements and site links.
As described in the following listing (cf. \cite[Figure 3]{wikidataRDF}), each statement consists of a claim and an optional reference. 
Each claim consists of a property - value pair, and optional qualifiers.  
\footnote{@prefix wkdt: $<http://wikidata.org/entity/>$ .}
\begin{lstlisting}[label={lst:wikidataexample}]
% Douglas Adams (Q42) spouse is Jane Belson (Q14623681)
%   - start time (P580) 25 November 1991, end time (P582) 11 May 2001. 
wkdt:Q42 wkdt:P26s wkdt:Q42Sb88670f8-456b-3ecb-cf3d-2bca2cf7371e.
wkdt:Q42Sb88670f8-456b-3ecb-cf3d-2bca2cf7371e wkdt:P580q wkdt:VT74cee544. 
wkdt:VT74cee544 rdf:type :TimeValue.;
 :time "1991-11-25"^^xsd:date;
 :timePrecision "11"^^xsd:int;   :preferredCalendar wkdt:Q1985727. 
wkdt:Q42Sb88670f8-456b-3ecb-cf3d-2bca2cf7371e wkdt:P582q wkdt:VT162aadcb.
wkdt:VT162aadcb rdf:type :TimeValue;
 :time "2001-5-11"^^xsd:date;
 :timePrecision "11"^^xsd:int;   :preferredCalendar wkdt:Q1985727.
\end{lstlisting}
%

\vspace{-1em}
\paragraph{\dbpedia} 
\cite{dbpedia-swj} The semantic extraction of information from Wikipedia is accomplished using the DBpedia Information Extraction Framework (DIEF).
The DIEF is able to process input data from several sources provided by Wikipedia.
The actual extraction is performed by a set of pluggable \emph{Extractors}, which rely on certain \emph{Parsers} for different data types.
Since 2011, DIEF is extended to provide better knowledge coverage for internationalized content~\cite{Kontokostas2012} and allows the easier integration of different Wikipedia language editions.


\vspace{-1em}
\section{Conversion Process}
\label{sec:ConversionProcess}

The \dbpedia Information Extraction Framework observed major changes to accommodate the extraction of data in \wikidata.
The major difference between \wikidata and the other Wikimedia projects \dbpedia extracts is that \wikidata uses JSON instead of WikiText to store items.

In addition to some \dbpedia provenance extractors that can be used in any MediaWiki export dump, we defined 10 additional \wikidata extractors to export as much knowledge as possible out of \wikidata.
These extractors can get labels, aliases, descriptions, different types of sitelinks, references,  statements and qualifiers. 
For statements we define a RawWikidataExtractor that extracts all available information but uses our reification scheme (cf. \autoref{sec:dataset}) and the \wikidata properties and the R2RWikidataExtractor that uses a mapping-based approach to map, in real-time, \wikidata statements to the \dbpedia ontology.

\vspace{-1em}
\subsubsection{Wikidata Property Mappings}
\label{sec:mappings}
In the same way the \dbpedia mappings wiki defines infobox to ontology mappings, in the context of this work we define \wikidata property to ontology mappings.
\wikidata property mappings can be defined both as \emph{Schema Mappings} and as \emph{Value Transformation Mappings}.

\vspace{-1em}
\paragraph{Schema Mappings}
The \dbpedia mappings wiki\footnote{\url{http://mappings.dbpedia.org}} is a community effort to map Wikipedia infoboxes to the \dbpedia ontology and at the same time crowd-source the \dbpedia ontology. 
Mappings between \dbpedia properties and \wikidata properties are expressed as \prop{owl:equivalentProperty} links in the property definition pages, e.g. \prop{dbo:birthPlace} is equivalent to \prop{wkdt:P569}.\footnote{\url{http://mappings.dbpedia.org/index.php/OntologyProperty:BirthDate}}
Although \wikidata does not define class in terms of \prop{RDFS} or \prop{OWL} we use \prop{OWL} punning to define \prop{owl:equivalentClass} links between the \dbpedia classes and the related \wikidata items, e.g. \prop{dbo:Person} is equivalent to \prop{wkdt:Q5}.\footnote{\url{http://mappings.dbpedia.org/index.php/OntologyClass:Person}}

\vspace{-1em}
\paragraph{Value Transformations}
At the time of writing, the value transformation takes the form of a JSON structure that binds a \wikidata property to one or more value transformation strings.
A complete list of the existing value transformation mappings can be found in the DIEF.
\footnote{\url{https://github.com/dbpedia/extraction-framework/blob/2ab6a15d8ecd5fc9dc6ef971a71a19ad4f608ff8/dump/config.json}}
The value transformation strings that may contain special placeholders in the form of a   `\textdollar{}' sign as functions.
If no `\textdollar{}' placeholder is found, the mapping is considered constant. e.g. \verb|"P625": {"rdf:type": "geo:SpatialThing"}|.
In addition to constant mappings, one can define the following functions:

\vspace{-1em}
\begin{description}
\item[\textdollar{1}] replaces the placeholder with the raw \wikidata value. e.g. 
\\\verb|  "P1566": {"owl:sameAs": "http://sws.geonames.org/$1/"}|.

\item[\textdollar{2}] replaces the placeholder with a space the wiki-title value, used when the value is a Wikipedia title and needs proper whitespace escaping. e.g. 
\\\verb|  "P154": {"logo": "http://commons.wikimedia.org/wiki/Special:FilePath/$2"},"|. 

\item[\textdollar{getDBpediaClass}] Using the schema class mappings, tries to map the current value to a \dbpedia class.
This function is used to extract \Verb|rdf:type| and \verb|rdfs:subClassOf| statement from the respective \wikidata properties. e.g
\\\verb|  "P31": {"rdf:type": "$getDBpediaClass"}|
\\\verb|  "P279":{"rdfs:subClassOf": "$getDBpediaClass"}|

\item[\textdollar{getLatitude}, \textdollar{getLongitude} \& \textdollar{getLongitude}] Geo-related functions to extract coordinates from values. Following is a complete geo mapping that the extracts geo coordinates similar to the \dbpedia coordinates dataset. For every occurrence of the property P625, four triples - one for every mapping - are generated:

\begin{minipage}{.48\textwidth}
\begin{lstlisting}[label={lst:geoconfiguration}]
"P625":[{"rdf:type":"geoSpatialThing"}, 
 {"geo:lat": "$getLatitude" },
 {"geo:long": "$getLongitude"},  
 {"georss:point": "$getGeoRss"}]
\end{lstlisting}
\end{minipage}
\begin{minipage}{.48\textwidth}
\begin{lstlisting}[label={lst:geoexample}]
DW:Q64 rdf:type geo:SpatialThing ;
 geo:lat "52.51667"^^xsd:float ;
 geo:long "13.38333"^^xsd:float ;
 geo:point "52.51667 13.38333" .
\end{lstlisting}
\end{minipage}
\end{description}

\vspace{-1em}
\paragraph{Mappings Application}
The R2RWikidataExtractor merges the schema \& value transformation property mappings and for every statement or qualifier it encounters, if mappings for the current \wikidata property exist, it tries to apply them and emit the mapped triples.

\vspace{-1.2em}
\subsubsection{Additions and Post Processing Steps}
\label{sec:post_processing}
Besides the basic extraction phase, additional processing steps are added in the workflow.

\vspace{-1em}
\paragraph{Type Inferencing}
In a similar way \dbpedia calculates transitive types for every resource, the \dbpedia Information Extraction Framework was extended to generate these triples directly at extraction time.
As soon as an \prop{rdf:type} triple is detected from the mappings, we try to identify the related \dbpedia class.
If a \dbpedia class is found, all super types are assigned to a resource.

\vspace{-1em}
\paragraph{Transitive Redirects}
\dbpedia has already scripts in place to identify, extract and resolve redirects.
After the redirects are extracted, a transitive redirect closure (excluding cycles) is calculated and applied in all generated datasets by replacing the redirected IRIs to the final ones. 

\vspace{-1em}
\paragraph{Validation}
The \dbpedia extraction framework already takes care of the correctness of the extracted datatypes during extraction.
We provide two additional steps of validation. The first step is performed in real-time during extraction and checks if the property mappings has a compatible \prop{rdfs:range} (literal or IRI) with the current value. The rejected triples are stored for feedback to the \dbpedia mapping community.
The second step is performed in a post-processing step and validates if the type of the object IRI is disjoint with the \prop{rdfs:range} of the property.
These errors, although they are excluded from the SPARQL endpoint and the Linked Data interface, are offered for download.

\begin{table}[t]
\scriptsize
\centering
    \begin{tabular}{ | l | r  | p{7cm} |}
    \hline
    \textbf{Title} & \textbf{Triples} & \textbf{Description} \\ \hline
    Provenance & 17,771,394 & PageIDs \& revisions \\ \hline
    Redirects & 434,094 & Explicit \& transitive redirects \\ \hline
    Aliases & 4,922,617 & Resource aliases with dbo:alias \\ \hline
    Labels & 61,668,295 & Labels with rdfs:label\\ \hline
    Descriptions & 95,667,863 & Descriptions with dbo:description \\ \hline 
    Sitelinks & 41,543,058 & \dbpedia inter-language links\\ \hline
    \wikidata links & 17,562,043 & Links to original \wikidata URIs\\ \hline
    Mapped facts & 90,882,327 & Aggregated mapped facts\\ \hline
    - Types & 8,579,745 & Direct types from the \dbpedia ontology \\ \hline
    - Transitive Types & 48,932,447 & Transitive types from the \dbpedia ontology \\ \hline
    - Coordinates & 6,415,120 & Geo coordinates \\ \hline
    - Images & 1,519,368 & Depictions using foaf:depiction \& dbo:thumbnail \\ \hline
    - mappings & 22,270,694 & \wikidata statements with \dbpedia ontology\\ \hline
    - External links & 3,164,953 & sameAs links to external databases\\ \hline
    Mapped facts (R) & 138,936,782 & Mapped statements reified (all)\\ \hline
    Mapped facts (RQ) & 626,648 & Mapped qualifiers \\ \hline
    Raw facts & 59,458,835 & Raw simple statements (not mapped)\\ \hline
    Raw facts (R) & 237,836,221 & Raw statements reified\\ \hline
    Raw facts (RQ) & 1,161,294 & Raw qualifiers \\ \hline
    References & 34,181,399 & Reified statements references with dbo:reference\\ \hline
    Mapping Errors & 2,711,114 & Facts from incorrect mappings\\ \hline
    Ontology Errors & 3,398 & Facts excluded due to ontology inconsistencies\\ \hline
    \end{tabular}
    \caption{Description of the \DW datasets. (R) stands for a reified dataset and (Q) for a qualifiers dataset}
    \label{tab:description}
\vspace{-20pt}
\end{table}

\vspace{-1.2em}
\subsubsection{IRI Schemes}
\label{sec:urischemes}
As mentioned earlier, we decided to generate the \rdf datasets under the \url{wikidata.dbpedia.org} domain.
For example, \prop{wkdt:Q42} will be transformed to \prop{dw:Q42}\footnote{\prop{@prefix dw: <http://wikidata.dbpedia.org/resource/>} .}.

\vspace{-1em}
\paragraph{Reification}
In contrast to \wikidata, simple \rdf reification was chosen for the representation of qualifiers.
This lead to a simpler design and further reuse of the \dbpedia properties.
The IRI schemes for the \prop{rdf:Statement} IRIs follow the same verbose approach from \dbpedia to make them easily writable manually by following a specific pattern.
When the value is an IRI (\wikidata Item) then for a subject IRI Qs, a property Px and a value IRI Qv the reified statement IRI has the form \prop{dw:Qs\_Px\_Qv}.
When the value is a Literal then for a subject IRI Qs, a property Px and a Literal value Lv the reified statement IRI has the form \prop{dw:Qs\_Px\_H(Lv,5)}, where H() is a hash function that takes as argument a string (Lv) and a number to limit the size of the returned hash (5). 
The use of the hash function in the case of literals guarantees the IRI uniqueness and the value `5' is safe enough to avoid collisions and keep it short at the same time.
The equivalent representation of the \wikidata example in \autoref{sec:background} is:
\footnote{\DW does not provide precision. Property definitions exist in the \dbpedia ontology}
\vspace{-0.5em}
\begin{lstlisting}[label={lst:reificationexample},multicols=2]
dw:Q42_P26_Q14623681 a rdf:Statement ;
 rdf:subject dw:Q42 ;
 rdf:predicate dbo:spouse ;
 rdf:object dw:Q14623681 ;
 dbo:startDate "1991-11-25"^^xsd:date ;
 dbo:endDate "2001-5-11"^^xsd:date ;
\end{lstlisting}


\vspace{-1.2em}
\section{Dataset Description}
\label{sec:dataset}
\vspace{-0.7em}
A statistical overview of the \DW dataset is provided in \autoref{tab:description}.
We extract provenance information, e.g.~the MediaWiki page and revision IDs as well as redirects.
Aliases labels and descriptions are extracted from the related \wikidata item section and are similar to the \rdf data \wikidata provides.
A difference to \wikidata are the properties we chose to associate aliases and description.

\wikidata sitelinks are processed to provide three datasets: 
1) \prop{owl:sameAs} links between \DW IRIs and \wikidata IRIs (e.g. \prop{dw:Q42} \prop{owl:sameAs} \prop{wkdt:Q42}),
2) \prop{owl:sameAs} links between \DW IRIs and sitelinks converted to \dbpedia IRIs (e.g. \prop{dw:Q42} \prop{owl:sameAs} \prop{db-en:Douglas\_Adams}) and
3) for every language in the mappings wiki we generate \prop{owl:sameAs} links to all other languages (e.g. \prop{db-en:Douglas\_Adams} \prop{owl:sameAs} \prop{db-de:Douglas\_Adams}).
The latter is used for the \dbpedia releases in order to provide links between the different \dbpedia language editions.

Mapped facts are generated from the \emph{\wikidata property mappings} (cf. \autoref{sec:mappings}).
Based on a combination of the predicate and object value of a triple they are split in different datasets.
Types, transitive types, geo coordinates,  depictions and external \prop{owl:sameAs} links are separated.
The rest of the mapped facts are in the \emph{mappings} dataset.
The reified mapped facts (R) contains all the mapped facts as reified statements and the mapped qualifiers for these statements (RQ) are provided separate (cf. \autoref{lst:reificationexample}).

Raw facts consist of three datasets that generate triples with \DW IRIs and the original \wikidata properties.
The first dataset (raw facts) provides triples for simple statements.
The same statements are reified in the second dataset (R) and in the third dataset (RQ) we provide qualifiers linked in the reified statements.
Example of the raw datasets can be seen in \autoref{lst:reificationexample} by replacing the \dbpedia properties with the original \wikidata properties. 
These datasets provide full coverage and, except from the reification design and different namespace, can be seen as equivalent with the \wikidata \rdf dumps.

\wikidata statement references are extracted in the \emph{references} dataset using the reified statement resource IRI as subject and the \prop{dbo:reference} property.
Finally, in the mapping and ontology errors datasets we provide triples rejected according to \autoref{sec:post_processing}.

\begin{table}[t]
\parbox{.45\linewidth}{
\scriptsize
\centering
\begin{tabular}{| l | r |}
\hline
\textbf{Class} & \textbf{Count}\\
\hline
dbo:Agent & 2,884,505 \\ \hline
dbo:Person & 2,777,306 \\ \hline
geo:spatialThing & 2,153,258 \\ \hline
dbo:TopicalConcept & 1,907,203 \\ \hline
dbo:Taxon & 1,906,747 \\ \hline
\end{tabular}
\caption{Top classes}
\label{tab:class_stats}
\vspace{-10pt}
}
\hfill
\parbox{.45\linewidth}{
\scriptsize
\centering
\begin{tabular}{| l | r |}
\hline
\textbf{Property} & \textbf{Count}  \\\hline
owl:sameAs & 232,890,848 \\ \hline
rdf:type & 145,407,453 \\ \hline
dbo:description & 95,667,863 \\ \hline
rdfs:label & 61,704,172\\ \hline
rdfs:seeAlso & 5,125,945 \\ \hline
\end{tabular}
\caption{Top properties}
\label{tab:prop_stats}
\vspace{-10pt}
}
\end{table}

\begin{table}[t]
\parbox{.45\linewidth}{
\scriptsize
\centering
\begin{tabular}{| l | r |}
\hline
\textbf{Property} & \textbf{Count}  \\\hline
dbo:date & 301,085\\ \hline
dbo:startDate & 158,947\\ \hline
geo:point & 108,526\\ \hline
dbo:endDate & 50,058\\ \hline
dbo:country & 33,698\\ \hline
\end{tabular}
\caption{Top mapped qualifiers}
\label{tab:qualifiers_stats}
\vspace{-20pt}
}
\hfill
\parbox{.45\linewidth}{
\scriptsize
\centering
\begin{tabular}{| l | r |}
\hline
\textbf{Property} & \textbf{Count}  \\\hline
wd:P31 & 13,070,656\\ \hline
wd:P17 & 3,051,166 \\ \hline
wd:P21 & 2,604,741 \\ \hline
wd:P131 & 2,372,237\\ \hline
wd:P625 & 2,167,100\\ \hline
\end{tabular}
\caption{Top properties in Wikidata}
\label{tab:error_stats}
\vspace{-20pt}
}
\end{table}

\vspace{-1.1em}
\section{DBpedia WikiData In Use}
\vspace{-0.7em}
\textbf{Statistics and Evaluation} 
The statistics we present are based on the \wikidata XML dump from March 2015. 
We managed to generate a total of 1B triples with 131,926,822 unique resources.
In \autoref{tab:description} we provide the number of triples per combined datasets.

\vspace{-1em}
\paragraph{Class \& property statistics} 
We provide the 5 most popular \DW classes in \autoref{tab:class_stats}.
We managed to extract a total of 6.5M typed Things with Agents and SpatialThing as the most frequent types.
The 5 most frequent mapped properties in simple statements are provided in \autoref{tab:prop_stats} and the most popular mapped properties in qualifiers in  \autoref{tab:qualifiers_stats}. 
\wikidata does not have a complete range of value types and date paoperties are the most frefuent at the moment.

\vspace{-1em}
\paragraph{Mapping statistics}
In total, 270 value transformation mappings were defined along with 163 \prop{owl:equivalentProperty} and 318 \prop{owl:equivalentClass} schema mappings. \wikidata has 1465 properties defined with a total of 60,119,911 occurrences.
With the existing mappings we covered 77.2\% of the occurrences.

\vspace{-1em}
\paragraph{Redirects}
In the current dataset we generated 434,094 redirects -- including transitive.
When these redirects were applied to the extracted datasets, they replaced 1,161,291 triples to the redirected destination.
The number of redirects in Wikidata is small compared to the project size but is is also a relatively new project.
As the project matures in time the number of redirects will increase and resolving them will have an impact on the resulting data.

\vspace{-1em}
\paragraph{Validation}
According to \autoref{tab:description}, a total of 2.7M errors originated from schema mappings and 3,398 triples did not pass the ontology validation (cf. \autoref{sec:post_processing}).


\vspace{-1.2em}
\subsubsection{Access and Sustainability}
This dataset will be part of the official \dbpedia knowledge infrastructure and be published through the regular releases of \dbpedia, along with the rest of the \dbpedia language editions.
The first \dbpedia release that will include this dataset is due on April - May 2015 (2015A).
\dbpedia is a pioneer in adopting and creating best practices for Linked Data and \rdf publishing.
Thus, being incorporated into the \dbpedia publishing workflow guarantees:
a) long-term availability through the \dbpedia Association and
b) agility in following best-practices as part of the \dbpedia Information Extraction Framework.
In addition to the regular and stable releases of \dbpedia we provide more frequent dataset updates from the project website.\footnote{\url{http://wikidata.dbpedia.org/downloads}}

Besides the stable dump availability we created \url{http://wikidata.dbpedia.org} for the provision of a Linked Data interface and a SPARQL Endpoint.
The dataset is registered in DataHub\footnote{\url{http://datahub.io/dataset/dbpedia-wikidata}} and provides machine readable metadata as void\footnote{\url{http://wikidata.dbpedia.org/downloads/void.ttl}} and DataID\footnote{\scriptsize\url{http://wikidata.dbpedia.org/downloads/20150330/dataid.ttl}} \cite{dataID2014}.
%
Since the project is now part of the official \dbpedia Information Extraction Framework, our dataset reuses the existing user and developer support infrastructure. 
\dbpedia has a general discussion and developer list as well as an issue tracker\footnote{\url{https://github.com/dbpedia/extraction-framework/issues}} for submitting bugs.

\vspace{-1.2em}
\subsubsection{Use Cases}

Although it is early to identify all possible use cases for \DW, our main motivation was a) ease of use, b) vertical integration with the existing \dbpedia infrastructure and c) data integration and fusion.
Following we list SPARQL query examples for simple and reified statements.
Since \dbpedia provides transitive types directly, queries where e.g. someone asks for all `places' in Germany can be formulated easier. Moreover, \prop{dbo:country} can be more intuitive than \prop{wkdt:P17c}.
Finally, the \dbpedia queries can, in most cases directly or with minor adjustments, run on all \dbpedia language endpoints.
When someone is working with reified statents, the \dbpedia IRIs encode all possible information to visually identify the resources and items involved (cf. \autoref{sec:urischemes}) in the statement while \wikidata uses a hash string. In addition, querying for reified statement in \wikidata needs to properly suffix the \wikidata property with \prop{c/s/q}:
\begin{lstlisting}[label={lst:sparqlsimple}]
#Queries with simple statement
select * WHERE {              | select * WHERE {
 ?place a dbo:Place ;         |  ?place wkdt:P31c/wkdt:P279c* wkdt:Q2221906 ;
        dbo:country dw:Q183.} |         wkdt:P17c wkdt:Q183. }        

#Queries with reified statements
select ?person where {                     | SELECT ?person WHERE {
 ?statementUri rdf:statement ?person ;     |  ?person wkdt:P26s ?spouse. 
               rdf:predicate dbo:spouse ;  |  ?spouse wkdt:P580q ?marriageValue.
               dbo:startDate ?date.        |  ?marriageValue wo:time ?date.
 FILTER (?date < "2000-01-01"^^xsd:date) } |  FILTER (?date < "2000-01-01"^^xsd:date)   }
\end{lstlisting}

An additional important use case is data integration.
Converting a dataset to a more used and well-known schema, it makes it easier to integrate the data.
The fact the datasets are split according to the information they contain makes data consumption easier when someone needs a specific subset, e.g. coordinates.
The \DW dataset is also planned to be used as an enrichment dataset on top of \dbpedia and fill in semi-structured data that are being moved to \wikidata.
It is also part of short-term plan to fuse all \dbpedia data into a single namespace and the \DW dataset will have a prominent role in this effort.

\vspace{-1em}
\section{Conclusions and Future Work}
\vspace{-0.5em}
We present an effort to provide an alternative \rdf representation of \wikidata.
Our work involved the creation of 10 new \dbpedia extractors, a Wikidata2DBpedia mapping language and additional post-processing \& validation steps.
With the current mapping status we managed to generate over 1 billion \rdf triples.
In the future we plan to extend the mapping coverage as well as extend the language with new mapping functions and more advanced mapping definitions.


\vspace{-1em}
\small
\bibliographystyle{abbrv}
\bibliography{literature}

\end{document}